\documentclass[fleqn,11pt,twoside]{article}
\usepackage{amsthm}
\usepackage[fleqn,intlimits]{amsmath}
\usepackage{graphicx}

\makeatletter

\newcommand{\Name}[1]{\begin{flushleft}
                       \LARGE \bf #1
                       \end{flushleft}\vspace{-3mm}}

\newcommand{\Author}[1]{\begin{flushleft}
                       \it #1 \end{flushleft}}

\newcommand{\Address}[1]{\begin{flushleft}
                       \it #1 \end{flushleft}}

\newcommand{\FirstPageHead}[5]{
\begin{flushleft}
\raisebox{8mm}[0pt][0pt]
{\footnotesize \sf
\parbox{150mm}{ \qquad
 #1 #2 #3 
#4\hfill {\sc #5}}}\vspace{-13mm}
\end{flushleft}}

\newcommand{\evenhead}{Author \ name}
\newcommand{\oddhead}{Article \ name}

\renewcommand{\@evenhead}{
\hspace*{-3pt}\raisebox{-15pt}[\headheight][0pt]{\vbox{\hbox to \textwidth
{\thepage \hfil \evenhead}\vskip4pt \hrule}}}
\renewcommand{\@oddhead}{
\hspace*{-3pt}\raisebox{-15pt}[\headheight][0pt]{\vbox{\hbox to \textwidth
{\oddhead \hfil \thepage}\vskip4pt\hrule}}}
\renewcommand{\@evenfoot}{}
\renewcommand{\@oddfoot}{}

\long\def\@makecaption#1#2{%
  \vskip\abovecaptionskip
  \sbox\@tempboxa{\small \textbf{#1.}\ \ #2}%
  \ifdim \wd\@tempboxa >\hsize
    {\small \textbf{#1.}\ \ #2}\par
  \else
    \global \@minipagefalse
    \hb@xt@\hsize{\hfil\box\@tempboxa\hfil}%
  \fi
  \vskip\belowcaptionskip}

\newcommand{\JNMPnumberwithin}[3][\arabic]{%
  \@ifundefined{c@#2}{\@nocounterr{#2}}{%
    \@ifundefined{c@#3}{\@nocnterr{#3}}{%
      \@addtoreset{#2}{#3}%
      \@xp\xdef\csname the#2\endcsname{%
        \@xp\@nx\csname the#3\endcsname .\@nx#1{#2}}}}%
}

\newcommand{\resetfootnoterule} {
  \renewcommand\footnoterule{%
  \kern-3\p@
  \hrule\@width.4\columnwidth
  \kern2.6\p@}
}

\renewcommand{\footnoterule}{}

\newcommand{\be}{\begin{equation}}
\newcommand{\ee}{\end{equation}}
\newcommand{\ba}{\hspace*{-5pt}\begin{array}}
\newcommand{\ea}{\end{array}}
\newcommand{\p}{\partial}

\makeatother

\numberwithin{equation}{section}
\theoremstyle{definition}

\renewcommand{\ba}{\begin{array}}
\renewcommand{\ea}{\end{array}}
\newcommand{\beg}{\begin{eqnarray}}
\newcommand{\eeq}{\end{eqnarray}}
\newcommand{\bg}{\begin{eqnarray*}}

\newcommand{\ed}{\end{eqnarray*}}
\newcommand{\n}{\newline\hfill}

\renewcommand{\p}{\partial} 
 
\newcommand{\notlhd}{\lhd\kern-.8em{/}\ } 
\newcommand{\notexist}{\ \exists\kern-.5em{\raise.1em\hbox{/}}\ }

\newcommand{\pde}[2]{\frac{\p #1}{\p #2}}

\newcommand{\inp}{{\mbox{\vbox{\hrule width0ex\hbox{\vrule
 height0ex\kern3.8pt
\vbox{\kern2.5pt}\kern3.8pt \vrule height1.6ex}
\hrule width1.6ex}}}}

\begin{document}

\renewcommand{\evenhead}{M Euler, N Euler and N Petersson}
\renewcommand{\oddhead}{Linearisable Hierarchies of Evolution Equations}


\thispagestyle{empty}

\begin{flushleft}
\footnotesize \sf
\end{flushleft}

\FirstPageHead{\ }{\ }{\ }
{
}{{\bf \tiny
{ }}}


\Name{Linearisable Hierarchies of Evolution Equations
in (1+1) Dimensions}

\label{euler-firstpage}

\Author{Marianna EULER, Norbert EULER and Niclas PETERSSON}

\Address{Department of Mathematics,  Lule\aa\ University of Technology, \\
SE-971 87 Lule\aa, Sweden\\
E-mails: Norbert@sm.luth.se, Marianna@sm.luth.se}


\begin{abstract}
\noindent
In our article \cite{eul_eul}, {\it ``A tree of linearisable second-order
evolution equations by generalised hodograph transformations}
[J. Nonlin. Math. Phys. {\bf 8} (2001),  342-362] we presented
a tree of linearisable ($C$-integrable) second-order evolution equations
in (1+1) dimensions. Expanding this result we report
here the complete set of recursion operators for this tree
and present several linearisable ($C$-integrable) hierarchies
in (1+1) dimensions.
\end{abstract}



\section{Introduction} 

In \cite{eul_eul} we presented a tree of
linearisable (that is, $C$-integrable) second-order evolution
equations which can be transformed to linear partial differential
equations.
The transformation under which the classification was performed
in \cite{eul_eul} is the so-called {\it $x$-generalised hodograph
transformation}
defined as follows:
\beg
\label{n_hodo}
_n\mbox{\bf H}:\left\{\ba{l}
\displaystyle{dX(x,t)
=f_1(x,u)dx+f_2(x,u,u_{x}, u_{xx},\ldots
, u_{x^{n-1}})dt}
\\[3mm]
dT(x,t)=dt \\[3mm]
U(X,T)=g(x),
\ea\right.
\eeq
with $n=2,3,\ldots$ and
\begin{gather}
\label{poincare}
u_{t}\pde{f_1 }{u}=
\pde{f_2}{x}+u_{x}\pde{f_2 }{u}+
u_{xx}\pde{f_2 }{u_{x}}
+\cdots +u_{x^{n}}\pde{f_2}{u_{x^{n-1}}}.
\end{gather}
Applying this transforms on an $(1+1)$-dimensional
$n$th-order autonomous evolution equation
\begin{gather}
\label{gen_Fi}
U_{T}=P\left(U, U_{X},U_{XX}, \ldots, U_{X^n}\right)
\end{gather}
leads in general to a 
$(1+1)$-dimensional
$n$th-order $x$-dependent evolution equation of the form
\begin{gather}
\label{gen_Fj}
u_{t}=Q(x, u, u_{x},u_{xx}, \ldots, u_{x^n}).
\end{gather}
Obviously $P$ and $Q$ are related via (\ref{n_hodo}) and (\ref{poincare}).

The prolongations of (\ref{n_hodo}) are
\begin{gather}
\label{prolong}
U_T=-\frac{f_2}{f_1}\frac{dg}{dx},\qquad
U_X=\frac{\dot g}{f_1}\\
U_{X^n}=\frac{1}{f_1}\left(D_x \left(\frac{1}{f_1}\right)\right)^{n-2}\,
D_x\left(\frac{\dot g}{f_1}\right),\qquad n=2,3,\ \ldots,\notag
\end{gather}
where $\dot g=dg/dx$ and $D_x$ is the total derivative operator
\bg
D_x=\pde{\ }{x}+u_x\pde{\ }{u}+u_{xx}\pde{\ }{u_x}+\cdots
\ed
with
\bg
(D_x a)^2=D_x(aD_x a),\quad (D_x a)^3=D_x(a D_x(aD_x a)), \ \ldots
\ed
Following (\ref{prolong}), 
the relation between $f_1$ and $f_2$ for (\ref{gen_Fi}) is
\begin{gather}
\label{gen_f1f2}
f_2(x,u,u_x,\ldots,u_{x^{n-1}})=\left.
-\frac{f_1}{\dot g}\left[P(U,U_X,\ldots U_{X^n})\right]\right|_{\Omega},
\end{gather}
where
\begin{gather}
\Omega=\left\{U=g(x),\ U_X=\frac{\dot g}{f_1},\ \ldots,\
U_{X^n}=\frac{1}{f_1}\left(D_x \left(\frac{1}{f_1}\right)\right)^{n-2}\,
D_x\left(\frac{\dot g}{f_1}\right)\right\}
\end{gather}
Equation (\ref{gen_Fj}) then follows from condition
(\ref{poincare}).

\strut\hfill

It should be pointed out that the $x$-generalised hodograph
transformation (\ref{n_hodo})
is a generalisation of
the {\it extended hodograph transformation} introduced in \cite{clark},
namely
\begin{gather}
X(x,t)=\int^{x} f(u(\xi, t))d\xi\notag\\
T(x,t)=t\\
U(X,T)=x.\notag
\end{gather}


For second-order evolution equations we showed \cite{eul_eul}
that

\strut\hfill

\noindent
{\it The most general $(1+1)$-dimensional
second-order evolution equation which may be
constructed to be linearisable in
\beg
\label{lin_auto2}
U_{T}=U_{XX}+\lambda_1 U_{X}
+\lambda_2 U,\qquad \lambda_1,\lambda_2\in
\Re
\eeq
via the $x$-generalised
hodograph transformation (\ref{n_hodo})
is necessarily of the form
\beg
\label{gen-nonauto2}
u_{t}=F_1(x, u)u_{xx}
+F_2(x, u)u_{x}
+F_3(x, u)u_{x}^2
+F_4(x, u).
\eeq
}

This led to sixteen linearisable second-order evolution
equations, eight of which are autonomous (by autonomous equations we
mean equations which do not depend explicitly on their independent
variables
$x$ and $t$). These equations are listed
in \cite{eul_eul} together with their linearising
transformations. Only one equation of this class is
{\it autohodograph invariant}, i.e., invariant under an $x$-generalised
hodograph transformation, namely the equation
\beg
\label{int_1}
u_{t}=h(u)u_{xx}+\left\{h\right\}_{u} u^2_{x},
\eeq
where $h\in C^2(\Re)$, $dh/du\neq 0$ and
\beg
\label{bracket}
\left\{h\right\}_{u}:=-\frac{1}{2}\frac{dh}{du}
+h\left(\frac{dh}{du}\right)^{-1}\frac{d^2 h}{du^2}.
\eeq


One can show the following:

\strut\hfill


\noindent
{\it The most general $n$th-order evolution equation which is 
linearisable in
\begin{gather}
\label{auto-linear}
U_{T}=\lambda_0 U + \sum_{l=1}^n\lambda_l U_{X^l},\qquad \lambda_j\in \Re
\end{gather}
by repeatedly applying the transformation (\ref{n_hodo}),
is necessarily of the form
\begin{gather}
\label{gen-form1}
u_{t}=\sum_{r=1}^n\sum_{k_1,k_2,\ldots k_r=0}^r
F_{k_1k_2\ldots k_r}(x,u) u_{x}^{k_1}u_{xx}^{k_2}\cdots
u_{x^r}^{k_r}+F_0(x,u),
\end{gather}
where $F_{k_1k_2\ldots k_r}\in C^2(\Re)$ are functions of
$x$ and $u$, and
\bg
\sum_{j=1}^r j k_j=r,\quad 
k_l\in\{0,1,\ldots ,n\},\qquad 1\leq l\leq r,\qquad 1\leq r\leq n.
\ed
}


\strut\hfill

The objective in the present paper is to construct higher-order
linearisable autonomous evolution equations by the use of the tree of
second-order linearisable equations gven in \cite{eul_eul}. 
This is achieved by calculating the recursion operators for the
second-order equations. The hierarchies obtained by the recursion
operators are all linearisable via the $x$-generalised hodograph
transformations and are of the general quasi-linear form
(\ref{gen-form1}). We do, however, restrict ourselves to the
autonomous case.
We furthermore write the equations in potential form and
use the pure hodograph transformation and corresponding recursion operators
to construct hierarchies of autonomous nonlinear
evolution equations and their linearising transformations.

\strut\hfill

We point out that the $x$-generalised hodograph transformation can be
applied only on autonomous evolution equations and may produce 
nonautonomous evolution equation of the same order. However, it is
easy to show that any nonautonomous evolution equation
which has been
generated by such a transformation can always be made autonomous.
This is achieved by writing the obtained equation in potential form
followed by the pure hodograph transformation.

\section{Second-order linearisable equations and their potential forms}

The following eight second-order evolution equations were constructed
to be linearisable via the $x$-generalised hodograph transformation
(\ref{n_hodo}) \cite{eul_eul}:
\begin{align}
&u_t=h_1u_{xx}+\{h_1\}_uu_x^2, &\dot h_1(u)&\neq 0\tag{I}\\
&u_t=h_2u_{xx}+\lambda h_2u_x+\{h_2\}_u u_x^2,& \dot h_2(u)&\neq 0,\
\lambda\neq 0 \tag{II}\\
&u_t=h_3u_{xx}+\{h_3\}_u u_x^2+2\lambda_2h^{3/2}\dot h_3^{-1}, &
\dot h_3(u)&\neq 0,\ \lambda_2\neq 0 \tag{III}\\
&u_t=u_{xx}+\lambda_4u_x+h_4^{-1}(\lambda_2-\dot
h_4)u_x^2+h_4, & h_4(u)&\neq 0,\ \lambda_2\neq 0
\tag{IV.1}\\
&u_t=u_{xx}+\lambda_4u_x-h_4^{-1}\dot h_4 u_x^2+h_4, & h_4(u)&\neq 0
\tag{IV.2}\\
&u_t=h_5u_{xx}+\left(\lambda
h_5-\lambda_2\lambda^{-1}\right)u_x+\{h_5\}_u u_x^2, &
\dot h_5(u)&\neq 0,\ \lambda\neq 0,\ \lambda_2\neq 0
\tag{V}\\
&u_t=u_{xx}+h_6u_x+\ddot h_6\dot h_6^{-1}u_x^2, &
h_6(u)&\neq 0
\tag{VI}\\
&u_t=h_7u_{xx}+\lambda_3u_x+\{h_7\}_u u_x^2, & \dot h_7(u)&\neq 0,\ 
\lambda_3\neq 0
\tag{VII}\\
&u_t=u_{xx}+\lambda_8u_x+h_8u_x^2& &\tag{VIII}
\end{align}
Here $h_j$ are arbitrary $C^2$-functions depending on $u$
(with the indicated restrictions). The bracket $\{h_j\}_u$
is defined by (\ref{bracket}).
All $\lambda$'s are arbitrary constants, unless otherwise
stated.
Here and in the rest of this paper the overdot on $h_j$
denotes differentiation with respect to $u$.

The equations (I - VIII) listed above may be linearised to
(\ref{lin_auto2}), i.e.,
\bg
U_T=U_{XX}+\lambda_1U_X+\lambda_2U-\lambda_3,\qquad \lambda_j\in\Re
\ed
by the following transformations, respectively \cite{eul_eul}:

\strut\hfill

\noindent
(Trans-I) ${\displaystyle \quad 
x=U(X,T),\quad dt=dT,\quad h_1(u)=U_X^2.
}$\n

\noindent
(Trans-II) ${\displaystyle \quad 
x=\frac{1}{\lambda}\ln|\lambda U|,\quad dt=dT,\quad
h_2(u)=\frac{1}{\lambda^2}\left(\frac{U_X}{U}\right)^2.
}$\n

\noindent
(Trans-III) ${\displaystyle \quad
x=\frac{2}{\lambda_2}\left(\frac{U_{X}}{U}\right),\quad dt=dT,\quad
h_3(u)=\frac{4}{\lambda_2^2}\left[\pde{\ }{X}
\left(\frac{U_{X}}{U}\right)\right]^2.
}$\n

\noindent
(Trans-IV.1) ${\displaystyle\quad \lambda_2\neq 0:\quad 
dx=dX,\quad dt=dT,\quad
\int^u\frac{1}{h_4(\xi)}d\xi=\frac{1}{\lambda_2}\ln|\lambda U|.
}$\n

\noindent
(Trans-IV.2) ${\displaystyle\quad\lambda_2=0: \quad 
dx=dX,\quad dt=dT,\quad
\frac{u_x}{h_4(u)}=-\frac{1}{\lambda_3}U.
}$\n

\noindent
(Trans-V) ${\displaystyle \quad 
x=\frac{1}{\lambda}\ln|\lambda U|,\quad dt=dT,\quad
h_5(u)=\frac{1}{\lambda^2}\left(\frac{U_X}{U}\right)^2.
}$\n

\noindent
(Trans-VI) ${\displaystyle \quad 
dx=dX,\quad dt=dT,\quad
h_6(u)=2\frac{U_X}{U}.
}$\n

\noindent
(Trans-VII) ${\displaystyle \quad
x=U,\quad dt=dT,\quad
h_7(u)=U_X^2.
}$\n

\noindent
(Trans-VIII) ${\displaystyle \quad
dx=dX,\quad dt=dT,\quad
\int^u\exp\left(\int^\xi h_8(\xi')d\xi'\right)d\xi=U_X.
}$\n

We now write equations (I) - (VIII) in potential form and
introduce new arbitrary functions $\phi_j(x)$, which then lead to
new linearisable autonomous evolution equations after transforming
the corresponding potential equations by the pure hodograph transformation.
The linearising transformations for the equations so obtained
result by composing the
potentials and hodograph tranformations with the corresponding
transformations (Trans-I)--(Trans-VIII) listed above. 

\strut\hfill

\noindent
{\bf Equation (I):} We give two possibilities
for writing (I) in potential form:\n
{\bf(I.i)}: Let
\begin{gather}
 v_x(x,t)=h_1^{-1/2}(u)+\phi_1(x)\notag\\
 v_t(x,t)=-\frac{1}{2}h_1^{-1/2}(u)\dot h_1(u) u_{x}-\lambda_1,\notag
\end{gather}
where $\phi_1$ is an arbitrary function of $x$ and $\lambda_1\in \Re$.
Then $v_{xt}=v_{tx}$
leads to (I). The potential equation takes the form
\begin{gather}
\label{pot_1}
v_t=\frac{v_{xx}-\phi_{1x}(x)}{(v_x-\phi_1(x))^2}-\lambda_1.
\end{gather}
Transforming (\ref{pot_1}) by the pure hodograph transformation
\begin{gather}
\label{pure}
v(x,t)=\chi,\quad t=\tau,\quad x=V(\chi,\tau)
\end{gather}
leads to the autonomous equation
\begin{gather}
\label{new_I}
V_\tau=\frac{V_{\chi\chi}+\phi_1'(V)V_\chi^3}{(1-\phi_1(V)V_\chi)^2}+\lambda_1
V_\chi,
\end{gather}
where $\phi'=d\phi_1/dV$. By the given change of variables
and (Trans-I), it follows that (\ref{new_I})
linearises to
\begin{gather}
U_T=U_{XX}+\lambda_1 U_X\notag
\end{gather}
by the transformation
\begin{gather}
\label{trans-I-new}
\displaystyle{
V(\chi,\tau)=U(X,T),\quad d\tau=dT,\quad
V_\chi^{-1}-\phi_1(V)=U_{X}^{-1}}.
\end{gather}

\noindent
{\bf(I.ii)}: Let
\begin{gather}
v_x(x,t)=xh_1^{-1/2}(u)+\phi_1(x)\notag\\
 v_t(x,t)=h_1^{1/2}(u)
-\frac{x}{2}h_1^{-1/2}(u)\dot h_1(u)u_x-\lambda_1.\notag
\end{gather}
Here $\phi_1$ is an arbitrary function of $x$ and $\lambda_1$
is an arbitrary constant. Clearly $v_{xt}=v_{tx}$
leads to (I). The potential equation takes the form
\begin{gather}
\label{pot_1ii}
v_t=\frac{v_{xx}-\phi_{1x}(x)}{(v_x-\phi_1(x))^2}x^2
-\lambda_1.
\end{gather}
Transforming (\ref{pot_1ii}) by the pure hodograph transformation
(\ref{pure})
leads to the autonomous equation
\begin{gather}
\label{new_Iii}
V_\tau=\frac{V_{\chi\chi}
+\phi_1'(V)V_\chi^3}{(1-\phi_1(V)V_\chi)^2}V
+\lambda_1V_\chi,
\end{gather}
where $\phi'=d\phi_1/dV$. By the given change of variables and
(Trans-I), it follows that (\ref{new_Iii})
linearises to
\begin{gather}
U_T=U_{XX}+\lambda_1 U_X\notag
\end{gather}
by the transformation
\begin{gather}
\displaystyle{
V(\chi,\tau)=U(X,T),\quad d\tau=dT,\quad
V_\chi^{-1}-\phi_1(V)=UU_{X}^{-1}
}.
\end{gather}

\strut\hfill

The same procedure can be followed for equations (II) - (VIII).
We list the results below:

\strut\hfill

\noindent
{\bf Equation (II):} Two cases are given.\n
{\bf (II.i):} 
Let
\begin{gather}
v_x(x,t)=h_2^{-1/2}(u)+\phi_2(x)\notag\\
 v_t(x,t)=
-\frac{1}{2}h_2^{-1/2}(u)\dot h_2(u)u_x-\lambda
 h_2^{1/2}(u)-\lambda_1,\quad \lambda\neq 0.\notag
\end{gather}
The potential equation is then
\begin{gather}
\label{pot_2i}
v_t=\frac{v_{xx}-\phi_{2x}(x)}{(v_x-\phi_2(x))^2}
-\frac{\lambda}{v_x-\phi_2}
-\lambda_1.
\end{gather}
By the pure hodograph transformation (\ref{pure}), (\ref{pot_2i}) leads to
\begin{gather}
\label{new_IIi}
V_\tau=\frac{V_{\chi\chi}
+\phi_2'(V)V_\chi^3}{(1-\phi_2(V)V_\chi)^2}
+\frac{\lambda V_\chi^2}{1-\phi_2(V)V_\chi}
+\lambda_1V_\chi,
\end{gather}
which linearises to
\begin{gather}
U_T=U_{XX}+\lambda_1 U_X\notag
\end{gather}
by the transformation
\begin{gather}
\label{trans-II-new}
\displaystyle{
V(\chi,\tau)=\frac{1}{\lambda}\ln |U(X,T)|,\quad d\tau=dT,\quad
V_\chi^{-1}-\phi_2(V)=\lambda U U_{X}^{-1}
}.
\end{gather}
\noindent
{\bf (II.ii):} 
Let
\begin{gather}
v_x(x,t)=e^{\lambda x}h_2^{-1/2}(u)+\phi_2(x)\notag\\
 v_t(x,t)=
-\frac{1}{2}e^{\lambda x}h_2^{-1/2}(u)\dot h_2(u)u_x-\lambda_1,\quad
\lambda\neq 0.\notag
\end{gather}
The potential equation is then
\begin{gather}
\label{pot_2ii}
v_t=e^{2\lambda x}\frac{v_{xx}-\phi_{2x}(x)}{(v_x-\phi_2(x))^2}
-\frac{\lambda e^{2\lambda x}}{v_x-\phi_2(x)}
-\lambda_1.
\end{gather}
By the pure hodograph transformation (\ref{pure}), (\ref{pot_2ii}) leads to
\begin{gather}
\label{new_IIii}
V_\tau=e^{2\lambda V}\frac{V_{\chi\chi}
+\phi_2'(V)V_\chi^3}{(1-\phi_2(V)V_\chi)^2}
+e^{2\lambda V}\frac{\lambda V_\chi^2}{1-\phi_2(V)V_\chi}
+\lambda_1V_\chi,
\end{gather}
which linearises to
\begin{gather}
U_T=U_{XX}+\lambda_1 U_X\notag
\end{gather}
by the transformation
\begin{gather}
\displaystyle{
V(\chi,\tau)=\frac{1}{\lambda}\ln |U(X,T)|,\quad d\tau=dT,\quad
V_\chi^{-1}-\phi_2(V)=\lambda U^2 U_{X}^{-1}}.
\end{gather}

\noindent
{\bf Equation (III):} Two cases are given.\n
{\bf (III.i):} 
Let
\begin{gather}
v_x(x,t)=h_3^{-1/2}(u)+\phi_3(x)\notag\\
 v_t(x,t)=
-\frac{1}{2}h_3^{-1/2}(u)\dot h_3(u)u_x-\lambda_2 x-\lambda_1,\quad
\lambda_2\neq 0.\notag
\end{gather}
The potential equation is then
\begin{gather}
\label{pot_IIIi}
v_t=\frac{v_{xx}-\phi_{3x}(x)}{(v_x-\phi_3(x))^2}
-\lambda_2 x
-\lambda_1.
\end{gather}
By the pure hodograph transformation (\ref{pure}), (\ref{pot_IIIi}) leads to
\begin{gather}
\label{new_IIIi}
V_\tau=\frac{V_{\chi\chi}
+\phi_3'(V)V_\chi^3}{(1-\phi_3(V)V_\chi)^2}
+\lambda_2VV_\chi
+\lambda_1V_\chi,
\end{gather}
which linearises to
\begin{gather}
U_T=U_{XX}+\lambda_1 U_X\notag
\end{gather}
by the transformation
\begin{gather}
\label{trans-III-new}
\displaystyle{
V(\chi,\tau)=\frac{2}{\lambda_2}\frac{U_X}{U},\quad d\tau=dT,\quad
V_\chi^{-1}-\phi_3(V)=\frac{\lambda_2}{2}
\left(\pde{\ }{X}\left(\frac{U_X}{U}\right)\right)^{-1}}.
\end{gather}

\noindent
{\bf (III.ii):} 
Let
\begin{gather}
v_x(x,t)=xh_3^{-1/2}(u)+\phi_3(x)\notag\\
 v_t(x,t)=h_3^{1/2}(u)
-\frac{1}{2}xh_3^{-1/2}(u)\dot h_3(u)u_x
-\frac{\lambda_2x^2}{2}
-\lambda_1,\quad \lambda\neq 0,\ \lambda_2\neq 0.\notag
\end{gather}
The potential equation then takes the form
\begin{gather}
\label{pot_IIIii}
v_t=\left(
\frac{v_{xx}-\phi_{3x}(x)}{(v_x-\phi_3(x))^2}\right)x^2
-\frac{\lambda_2 x^2}{2}
-\lambda_1.
\end{gather}
By the pure hodograph transformation (\ref{pure}),
(\ref{pot_IIIii}) leads to
\begin{gather}
\label{new_IIIii}
V_\tau=V^2\left(
\frac{V_{\chi\chi}
+\phi_3'(V)V_\chi^3}{(1-\phi_3(V)V_\chi)^2}\right)
+\frac{\lambda_2}{2}V^2V_\chi
+\lambda_1V_\chi,
\end{gather}
which linearises to
\begin{gather}
U_T=U_{XX}+\lambda_1 U_X +\lambda_2 U\notag
\end{gather}
by the transformation
\begin{gather}
\displaystyle{
V(\chi,\tau)=\frac{2}{\lambda_2}\frac{U_X}{U},\quad d\tau=dT,\quad
V_\chi^{-1}-\phi_3(V)=\frac{U_X}{U}
\left(\pde{\ }{X}\left(\frac{U_X}{U}\right)\right)^{-1}}.
\end{gather}

\noindent
{\bf Equation (IV.1):} 
Let
\begin{gather}
v_x(x,t)=\exp\left(\lambda_2\int^u\frac{1}{h_4(\xi)}d\xi+rx\right)
+\phi_4(x)\notag\\
v_t(x,t)=\left(\frac{\lambda_2}{h_4(u)}u_x+\frac{\lambda_2}{r}\right)\exp
\left(\lambda_2\int^u\frac{1}{h_4(\xi)}d\xi+rx\right)-\lambda_1,\qquad
\lambda_2\neq 0,\notag
\end{gather}
where
\begin{gather}
\label{r}
r=\frac{\lambda_1}{2}\pm\left(\frac{\lambda_1^2}{4}-\lambda_2\right)^{1/2}.
\end{gather}
The potential equation then becomes
\begin{gather}
\label{pot_IV.1}
v_t=v_{xx}-\phi_{4x}(x)
+(\lambda_1-2r)(v_x-\phi_4(x))-\lambda_1.
\end{gather}
By the pure hodograph transformation (\ref{pure}),
(\ref{pot_IV.1}) leads to
\begin{gather}
\label{new_IV.1}
V_\tau=V_\chi^{-2}V_{\chi\chi}
+\phi_4'(V)V_\chi+(2r-\lambda_1)(1-\phi_4(V)
V_\chi)+\lambda_1 V_\chi
\end{gather}
which linearises to
\begin{gather}
U_T=U_{XX}+\lambda_1 U_X+\lambda_2 U\notag
\end{gather}
by the transformation
\begin{gather}
\label{trans-IV.1-new}
\displaystyle{
V(\chi,\tau)=X,\quad d\tau=dT,\quad
V_\chi^{-1}-\phi_4(V)=e^{r X} U},
\end{gather}
where $r$ is given by (\ref{r}).

\strut\hfill

\noindent
{\bf Equation (IV.2):} 
Let
\begin{gather}
v_x(x,t)=e^{\lambda_1 x}\int^u\frac{1}{h_4(\xi)}d\xi
+\phi_4(x)\notag\\
v_t(x,t)=e^{\lambda_1x}\frac{u_x}{h_4(u)}
+\frac{1}{\lambda_1}e^{\lambda_1x}
-\lambda_1,\quad \lambda_1\neq 0.\notag
\end{gather}
The potential equation then becomes
\begin{gather}
\label{pot_IV}
v_t=v_{xx}-\lambda_1(v_x-\phi_4(x)-\phi_{4x}(x))
+\frac{1}{\lambda_1}e^{\lambda_1x}-\lambda_1.
\end{gather}
By the pure hodograph transformation (\ref{pure}),
(\ref{pot_IV}) leads to
\begin{gather}
\label{new_IV.2}
V_\tau=V_\chi^{-2}V_{\chi\chi}
+\phi_4'(V)V_\chi+\lambda_1(1-\phi_4(V)
V_\chi)-\frac{1}{\lambda_1}e^{\lambda_1 V}V_\chi+\lambda_1 V_\chi
\end{gather}
which linearises to
\begin{gather}
U_T=U_{XX}+\lambda_1 U_X\notag
\end{gather}
by the transformation
\begin{gather}
\label{trans-IV.2-new}
\displaystyle{
V(\chi,\tau)=X,\quad d\tau=dT,\quad
V_\chi^{-1}-\phi_4(V)=e^{\lambda_1 X}
\int^X U(\xi,T)d\xi}.
\end{gather}

\noindent
{\bf Equation (V):} 
Let
\begin{gather}
v_x(x,t)=h_5^{-1/2}(u)+\phi_5(x)\notag\\
v_t(x,t)=-\frac{1}{2}h_5^{-1/2}\dot h_5(u)u_x-\lambda h_5^{1/2}(u)
-\frac{\lambda_2}{\lambda}h_5^{-1/2}(u)-\lambda_1,\quad \lambda\neq 0.
\notag
\end{gather}
The potential equation then becomes
\begin{gather}
\label{pot_V}
v_t=\frac{v_{xx}-\phi_{5x}(x)}{(v_x-\phi_5(x))^2}
-\frac{\lambda}{v_x-\phi_5(x)} 
-\frac{\lambda_2}{\lambda}(v_x-\phi_5(x))-\lambda_1.
\end{gather}
By the pure hodograph transformation (\ref{pure}),
(\ref{pot_V}) leads to
\begin{gather}
\label{new_V}
V_\tau=\frac{V_{\chi\chi}+\phi_5'(V)V_\chi^3}{(1-\phi_5(V)V_\chi)^2}
+\frac{\lambda V_\chi^2}{1-\phi_5(V)V_\chi}
+\frac{\lambda_2}{\lambda}(1-\phi_5(V)V_\chi)+\lambda_1V\chi
\end{gather}
which linearises to
\begin{gather}
U_T=U_{XX}+\lambda_1 U_X\notag+\lambda_2 U
\end{gather}
by the transformation
\begin{gather}
\label{trans-V-new}
\displaystyle{
V(\chi,\tau)=\frac{1}{\lambda}\ln|\lambda U(X,T)|,
\quad d\tau=dT,\quad
V_\chi^{-1}-\phi_5(V)=\lambda\frac{U}{U_X}.
}
\end{gather}

\noindent
{\bf Equation (VI):} 
Let
\begin{gather}
v_x(x,t)=h_6(u)
+\phi_6(x)\notag\\
v_t(x,t)=\dot h_6(u)u_x+\frac{1}{2}h_6^2(u).\notag
\end{gather}
The potential equation then takes the form
\begin{gather}
\label{pot_VI}
v_t=v_{xx}-\phi_6(x)+\frac{1}{2}(v_x-\phi_6(x))^2.
\end{gather}
By the pure hodograph transformation (\ref{pure}), (\ref{pot_VI}) leads to
\begin{gather}
\label{new_VI}
V_\tau=V_\chi^{-2}V_{\chi\chi}
+\phi_6'(V)V_\chi-\frac{1}{2}(1-\phi_6(V)V_\chi)^2V_\chi^{-1}
\end{gather}
which linearises to
\begin{gather}
U_T=U_{XX}
\end{gather}
by the transformation
\begin{gather}
\label{trans-VI-new}
\displaystyle{
V(\chi,\tau)=X,\quad d\tau=dT,\quad
V_\chi^{-1}-\phi_6(V)=2\frac{U_X}{U}.}
\end{gather}

\noindent
{\bf Equation (VII):} 
Let
\begin{gather}
v_x(x,t)=h_7^{-1/2}(u)+\phi_7(x)\notag\\
v_t(x,t)=-\frac{1}{2}h_7^{-1/2}\dot h_7(u)u_x-\lambda_3 h_7^{-1/2}(u)
-\lambda_1.
\notag
\end{gather}
The potential equation then becomes
\begin{gather}
\label{pot_VII}
v_t=\frac{v_{xx}-\phi_{7x}(x)}{(v_x-\phi_7(x))^2}
-\lambda_3(v_x-\phi_7(x))
-\lambda_1.
\end{gather}
By the pure hodograph transformation (\ref{pure}), (\ref{pot_VII}) leads to
\begin{gather}
\label{new_VII}
V_\tau=\frac{V_{\chi\chi}+\phi_7'(V)V_\chi^3}{(1-\phi_7(V)V_\chi)^2}
+\lambda_3(1-\phi_7(V)V_\chi)+\lambda_1V\chi
\end{gather}
which linearises to
\begin{gather}
U_T=U_{XX}+\lambda_1 U_X\notag+\lambda_3
\end{gather}
by the transformation
\begin{gather}
\label{trans-VII-new}
\displaystyle{
V(\chi,\tau)=U(X,T),
\quad d\tau=dT,\quad
V_\chi^{-1}-\phi_7(V)=U_X^{-1}.
}
\end{gather}

\noindent
{\bf Equation (VIII):} 
Let
\begin{gather}
v_x(x,t)=e^{\lambda_1 x}\int^u\exp\left[ \int^\xi
h_8(\xi')d\xi'\right]d\xi+\phi_8(x)
\notag\\
v_t(x,t)=e^{\lambda_1x}\exp\left[ \int^u
h_8(\xi)d\xi\right]\,u_x-\lambda_1.\notag
\end{gather}
The potential equation then takes the form
\begin{gather}
\label{pot_VIII}
v_t=v_{xx}-\lambda_1(v_x-\phi_8(x))
-\phi_{8x}(x)-\lambda_1
\end{gather}
By the pure hodograph transformation (\ref{pot_VIII}) leads to
\begin{gather}
\label{new_VIII}
V_\tau=V_\chi^{-2}V_{\chi\chi}
+\phi_8'(V)V_\chi+\lambda_1(1-\phi_8(V)
V_\chi)+\lambda_1 V_\chi
\end{gather}
which linearises to
\begin{gather}
U_T=U_{XX}+\lambda_1 U_X\notag
\end{gather}
by the transformation
\begin{gather}
\label{trans-VIII-new}
\displaystyle{
V(\chi,\tau)=X,\quad d\tau=dT,\quad
V_\chi^{-1}-\phi_8(V)=e^{\lambda_1 X}U_X}.
\end{gather}

\strut\hfill

The same procedure could now, in principle, be applied again
on these new equations (\ref{new_I}), (\ref{new_Iii}),
(\ref{new_IIi}), (\ref{new_IIii}), (\ref{new_IIIi}),
(\ref{new_IIIii}), (\ref{new_IV.1}), (\ref{new_IV.2}), (\ref{new_V}),
(\ref{new_VI}), (\ref{new_VII}), and (\ref{new_VII}),
to construct new chains of linearisable equations. We present here
only one example.

\strut\hfill

\noindent
{\bf An example to further extend (\ref{new_I})}:
Let
\begin{gather}
W_\chi(\chi,\tau) = V+\Omega(\chi)\notag\\
W_\tau(\chi,\tau) = \frac{V\chi}{(1-\phi_1(V) V_\chi)},\notag
\end{gather}
where $\Omega$ is an arbitrary function of $\chi$. We set
$\lambda_1=0$
in (\ref{new_I}). The potential equation
then takes the form
\begin{gather}
\label{pot_Ii_next}
W_\tau=\frac{W_{\chi\chi}-\Omega_\chi}{1-(W_{\chi\chi}-\Omega_\chi)
\phi_1(W_\chi-\Omega)}.
\end{gather}
Performing the pure hodograph transformation
\begin{gather}
W(\chi,\tau)=\xi,\quad \chi=\omega(\xi,\eta),\quad
\tau=\eta\notag
\end{gather}
on (\ref{pot_Ii_next}) leads to
\begin{gather}
\label{new_next_Ii}
\omega_\eta=
\frac{\omega_\xi\omega_{\xi\xi}+\omega_\xi^4\Omega_\omega}{
\omega_\xi^3+(\omega_{\xi\xi}+\omega_\xi^3\Omega_\omega)
\phi_1(\omega_\xi^{-1}-\Omega)}
\end{gather}
which may be linearised to
\bg
U_X=U_{XX}
\ed
by the transformation
\begin{gather}
\omega_\xi^{-1}-\Omega(\omega)=U(X,T)\notag\\
d\eta=dT\\
\omega_\xi^{-3}\omega_{\xi\xi}+\Omega'(\omega)
=\frac{U_X}{U_X\phi_1(U)-1}\notag
\end{gather}
where $\Omega'=d\Omega/d\omega$.

A detailed analysis of this type of extensions will be considered
elsewhere.


\section{Recursion operators and hierarchies of linearisable equations}

In this section we give the recursion operators for the linearisable
equations listed in Section 2. Before we list our results, we
mention some relevant facts concerning recursion operators of
evolution equations.

A recursion operator in two independent variables $x$ and $t$
is a linear integro-differential operator
of the form
\begin{gather}
R[u]=\sum_{j=0}^lP_jD_x^j+\sum_{j=1}^sQ_jD_x^{-j},
\end{gather}
where $D_x$ denotes the total $x$-derivative and
$D_x^{-j}$ the $j$-fold product of the inverse of $D_x$.
The coefficients $P_j$ and $Q_j$ depend in general on $x$, $t$ and
$u$, as well as a finite number of
$x$ derivatives of $u$. These operators were introduced by Olver
\cite{Olver77} to generate (infinite) sequences of Lie-B\"acklund (also
called Generalised) symmetry generators. For an $n$th-order evolution
equation in $u$,
\begin{gather}
\label{n-th-EE}
u_t=F(x,t,u,u_x,u_{xx},\ldots,u_{x^n}),
\end{gather}
a Lie-B\"acklund symmetry generator $Z$ is of the form
\begin{gather}
Z=\eta(x,t,u,u_x,u_{xx},\ldots,u_{x^q})\pde{\ }{u},\qquad q>n,
\end{gather}
if it exists.
The recursion operator is such that when acting on the Lie-B\"acklund
symmetry generator,
the result is still a Lie-B\"acklund symmetry of the same evolution
equation (\ref{n-th-EE}), i.e.,
\begin{gather}
\eta_{i+1}=R[u]\eta_i.
\end{gather}
The recursion operator then satisfy the commutation relation
\begin{gather}
[L[u]\, ,\, R[u]]=D_tR[u],
\end{gather}
where $L[u]$ is the linear operator
\begin{gather}
L[u]=\pde{F}{u}+\pde{F}{u_x}D_x+\pde{F}{u_{xx}}D_x^2+\cdots +
\pde{F}{u_{x^n}}D_x^n
\end{gather}
and $D_tR[u]$ calculates the explicit derivative with respect to $t$.
For more details we refer to \cite{Olver, Bluman, fokas87}.

In this paper we consider only autonomous evolution equations, i.e.,
equations invariant under translation in
$t$ and $x$. Therefore the equations admit the point symmetry generators
\begin{gather}
u_t\pde{\ }{u},\qquad u_x\pde{\ }{u}.
\end{gather}
A hierarchy of evolution equations can
then be obtained by applying the recursion operator on the
$t$-translation symmetry, or equivalently on $F$, i.e.,
\begin{gather}
u_t=R^m[u]F,\qquad m\in {\cal N}.
\end{gather}

\subsection{Recursion operators for (I) - (VIII)}

Below we list the recursion operators for the eight linearisable
equations (I - VIII) listed in Section 2.
Note that equations (II) and (V) share the same recursion operator,
as do equations (I) and (VII):
\bg
&\mathbf{I} & R_1[u]=h_1^{1/2}D_{x}+\dfrac{\{h_1\}_u}{h_1^{1/2}}u_{x}
+\dfrac{1}{2}
\left( h_1u_{xx}+\{h_1\}_uu_{x}^{2}\right)
D_{x}^{-1}\dfrac{\dot{h_1}}{h_1^{3/2}} \\ 
& &  \\ 
&\mathbf{II} & R_2[u]=h_2^{1/2}D_{x}+\dfrac{\{h_2\}_u}{h_2^{1/2}}u_{x}
+\lambda h_2^{1/2}+
\dfrac{1}{2}\left( h_2u_{xx}+\{h_2\}_uu_{x}^{2}
+\lambda h_2u_{x}\right) D_{x}^{-1}
\dfrac{\dot{h_2}}{h_2^{3/2}}
\\ 
& & \\ 
&\mathbf{III} & R_3[u]=h_3^{1/2}D_{x}+\dfrac{\{h_3\}_u}{h_3^{1/2}}u_{x}
+\dfrac{1}{2}
\lambda _{2}x+\dfrac{1}{2}\left( h_3u_{xx}+\{h_3\}_uu_{x}^{2}+2\lambda _{2}
\dfrac{h_3^{3/2}}{\dot{h_3}}\right)
D_{x}^{-1}\dfrac{\dot{h_3}}{h_3^{3/2}}\\
& &  \\ 
& \mathbf{IV} & R_4[u]=D_{x}+\dfrac{1}{h_4}\left( \lambda _{2}-\dot{h_4}\right)
u_{x} \\ 
& & \\ 
& \mathbf{V} & R_5[u]=h_5^{1/2}D_{x}+\dfrac{\{h_5\}_u}{h_5^{1/2}}u_{x}+\lambda
h_5^{1/2}
+\dfrac{1}{2}\left( h_5u_{xx}+\{h_5\}_u u_{x}^{2}
+\lambda h_5u_{x}\right) D_{x}^{-1}
\dfrac{\dot{h_5}}{h_5^{3/2}} \\ 
& &  \\ 
& \mathbf{VI} & R_6[u]=D_{x}+\dfrac{\ddot{h_6}}{\dot{h_6}}u_{x}
+\dfrac{1}{2}h_6+
\dfrac{1}{2}u_{x}D_{x}^{-1}\dot{h_6} \\ 
& & \\ 
&\mathbf{VII} & R_7[u]=h_7^{1/2}D_{x}+\dfrac{\{h_7\}_u}{h_7^{1/2}}u_{x}
+\dfrac{1}{2}
\left( h_7u_{xx}+\{h_7\}u_{x}^{2}\right)
D_{x}^{-1}\dfrac{\dot{h_7}}{h_7^{3/2}} \\ 
& & \\ 
&\mathbf{VIII} & R_8[u]=D_{x}+h_8u_{x}
\ed
The recursion operators $R_4$ and $R_6$ are also given in \cite{fokas80}. 

Below we list the first nontrivial Lie-B\"{a}cklund symmetries of the eight
linearisable second-order evolution equations (I - VIII), with the same
equation numbers. We note that all
symmetries are of order three and only equation (III) has an $x$-dependent
third-order symmetry. 

\bg
&\mathbf{I} & Z_1=h_1^{3/2}
\left(u_{xxx}+3\dfrac{\ddot{h_1}}{\dot{h_1}}u_{x}u_{xx}+
\dfrac{\dddot{h_1}}{\dot{h_1}}u_{x}^{3}\right)
\dfrac{\partial }{\partial u} \\ & & \\ 
& \mathbf{II} & Z_2=h_2^{3/2}\left( u_{xxx}
+3\dfrac{\ddot{h_2}}{\dot{h_2}}u_{x}u_{xx}+
\dfrac{\dddot{h_2}}{\dot{h_2}}u_{x}^{3}
+3\lambda u_{xx}+3\lambda \dfrac{\ddot{h_2}}{\dot{h_2}}u_{x}^{2}
+2\lambda ^{2}u_{x}\right) \dfrac{\partial }{\partial u}
\\ 
& & \\ 
&\mathbf{III} & Z_3=\left\{ h_3^{3/2}\left( u_{xxx}
+3\dfrac{\ddot{h_3}}{\dot{h_3}}
u_{x}u_{xx}+\dfrac{\dddot{h_3}}{\dot{h_3}}u_{x}^{3}\right)
\right.\\
& & \\
& & \qquad +\dfrac{3}{2}\lambda_{2}x\left( h_3u_{xx}+\{h_3\}u_{x}^{2}
+2\lambda _{2}\dfrac{h_3^{3/2}}{\dot{h_3}}\right)
 \left.\vphantom{\dfrac{\ddot{h_3}}{\dot{h_3}}}
+3\lambda _{2}h_3u_{x}\right\} \dfrac{\partial }{\partial u}
\ed
\bg
&\mathbf{IV} & Z_4=\left\{ u_{xxx}+3\dfrac{1}{h_4}\left( \lambda _{2}-\dot{h_4}
\right) u_{x}u_{xx}\right.\\
& & \\
& & \quad\ \left.\vphantom{\dfrac{\ddot{h_3}}{\dot{h_3}}}
+\dfrac{1}{h_4^{2}}\left( \lambda _{2}^{2}-3\lambda _{2}
\dot{h_4}+2\dot{h_4}^{2}-h\ddot{h_4}\right) u_{x}^{3}
+\lambda _{2}u_{x}\right\} 
\dfrac{\partial }{\partial u} \\ 
& & \\ 
&\mathbf{V} & Z_5=h^{3/2}\left( u_{xxx}
+3\dfrac{\ddot{h_5}}{\dot{h_5}}u_{x}u_{xx}+
\dfrac{\dddot{h_5}}{\dot{h_5}}u_{x}^{3}+3\lambda u_{xx}
+3\lambda \dfrac{\ddot{h_5}}{\dot{h_5}}u_{x}^{2}
+2\lambda^{2}u_{x}\right) \dfrac{\partial }{\partial u}\\ 
& & \\ 
&\mathbf{VI} & Z_6=\left\{
u_{xxx}+3\dfrac{\ddot{h_6}}{\dot{h_6}}u_{x}u_{xx}
+\dfrac{\dddot{h_6}}{\dot{h_6}}u_{x}^{3}
+\dfrac{3}{2}h_6u_{xx}\right.\\
& &\\
& & \quad\ \left.\vphantom{\dfrac{\ddot{h_3}}{\dot{h_3}}}
+\dfrac{3}{2}\left( \dot{h_6}+h_6
\dfrac{\ddot{h_6}}{\dot{h_6}}\right) u_{x}^{2}
+\dfrac{3}{4}h_6^{2}u_{x}\right\} 
\dfrac{\partial }{\partial u}
\ed
\bg
&\mathbf{VII} & Z_7=h_7^{3/2}\left( u_{xxx}
+3\dfrac{\ddot{h_7}}{\dot{h_7}}u_{x}u_{xx}+
\dfrac{\dddot{h_7}}{\dot{h_7}}u_{x}^{3}\right)
\dfrac{\partial }{\partial u} \\ 
& & \\ 
&\mathbf{VIII} & Z_8=\left\{ u_{xxx}+3h_8u_{x}u_{xx}
+\left( \dot{h_8}+h_8^{2}\right)
u_{x}^{3}\right\} \dfrac{\partial }{\partial u}
\ed

The hierarchies resulting from the second-order linearisable equations
(I) - (VIII)
can then be written in the form
\begin{gather}
\label{hier}
u_t=\left(R_k^m[u]+\sum_{i=0}^{m-1} R_k^{i}[u]C_{i}\right)F_k(u,u_x,u_{xx})
\end{gather}
where $m=1,2,\ldots$, $C_i$ are arbitrary constants, and $F_k$
denotes the r.h.s of the second-order equations (I) - (VIII).
The transformations linearising the hierarchies (\ref{hier})
are the same as the transformation (Trans-I) -- (Trans-VIII)
for the second-order equations (I) - (VIII) listed in Section 2,
whereby the linearised equation in $U$ with independent variables
$X$ and $T$ are of the same order as the equation from the hierarchy
which is being linearised. We'll refer to these linearisable hierarchies
as hierarchies (I) - (VIII), corresponding to the second-order
equations (I) - (VIII).

\strut\hfill

As an example we write explicitly the third-order
equation from the hierarchy (I) with $h_1=u^n$ ($n\in
{\cal Q}\backslash\{0\}$) using its recursion operator
$R_1$, which now takes the form
\begin{gather}
R_1[u]=u^{n/2}D_x+\frac{n-2}{2}u^{(n/2)-1}u_x
+\frac{n}{2}\left(u^n u_{xx}+\frac{n-2}{2}u^{n-1}u_x^2
 \right)D_x^{-1}u^{-(n/2)-1}.
\end{gather}
Thus
\bg
u_t=\left(R_1[u]+C_0\right)\left(u^n u_{xx}+\frac{n-2}{2}u^{n-1}u_x^2
  \right),
\ed
is the third-order linearisable equation, namely 
\begin{gather}
u_t=u^{3n/2}u_{xxx}
+3(n-1)u^{(3n/2)-1}u_xu_{xx}
+  C_0  u^nu_{xx}\notag\\
\qquad +\frac{C_0}{2}(n-2) u_x^2u^{n-1}
+(n-1)(n-2)u^{(3n/2)+1}u_x^3,
\end{gather}
which linearises to
\begin{gather}
U_T=U_{XXX}+C_0U_{XX}\notag
\end{gather}
by the transformation (Trans-I).

\subsection{Some examples of hierarchies (I)-(VIII) in the literature}
\noindent
{\bf Third-Order equations:}

\strut\hfill

\noindent
We give two equations from the list in \cite{calogero91}
by F Calogero:

\strut\hfill

\noindent
(3.138 \cite{calogero91}) ${\displaystyle \quad 
u_t=u^3u_{xxx}+(2u^3+au^2)u_x
+(3u^3+au^2)u_{xx}+3u^2u_x^2 + 3u^2u_xu_{xx},}$

\strut\hfill

\noindent
which is part of hierarchy $\mathbf{II}$, namely
\bg
u_t=\left(R_2[u]+C_0\right)\left(h_2u_{xx}+\lambda h_2 u_x+\{h_2\}_u u_x^2 \right),
\ed
with $h_2=u^2$, $\lambda=1$ and $C_0=a$ (arbitrary constant).
This equation linearises to
\begin{gather}
U_T=U_{XXX}+aU_{XX}\notag
\end{gather}
by the transformation
\begin{gather}
x=\ln U,\quad
dt=dT,\quad
u(x,t)=U^{-1}U_X.\notag
\end{gather}

\strut\hfill

\noindent
(3.133 \cite{calogero91}) ${\displaystyle \quad 
u_t=u_{xxx}+au_{xx}+3u_xu_{xx}+au_x^2+u_x^3,}$

\strut\hfill

\noindent
which is part of hierarchy $\mathbf{VIII}$, namely
\begin{gather}
u_t=\left(R_8[u]+C_0\right)\left(u_{xx}+h_8 u_x^2 \right),\notag
\end{gather}
with $h_8=1$ and $C_0=a$ (arbitrary constant).
This equation linearises to
\begin{gather}
U_T=U_{XXX}+aU_{XX}\notag
\end{gather}
by the transformation
\begin{gather}
dx=dX,\quad
dt=dT,\quad
u(x,t)=\ln\left|U_X\right|.\notag
\end{gather}

\strut\hfill

\noindent
An equations introduced by S Kawamoto \cite{kawamoto}:
\begin{gather}
u_t=u^3u_{xxx}+(1+\alpha)u^2u_{xx}\notag
\end{gather}
which is part of hierarchy $\mathbf{I}$, namely
\begin{gather}
u_t=\left(R_1[u]+C_0\right)\left(h_1u_{xx}+\{h_1\}_u u_x^2 \right),\notag
\end{gather}
with $h_1=u^2$, $\alpha=2$ and $C_0=0$. 
This equation linearises to
\begin{gather}
U_T=U_{XXX}\notag
\end{gather}
by the transformation
\begin{gather}
x=U,\quad
dt=dT,\quad
u(x,t)=U_X.\notag
\end{gather}

\noindent
{\bf Fourth-order equations:}

\strut\hfill

\noindent
An equation introduced by S Kawamoto \cite{kawamoto}:
\begin{gather}
u_t=u^4u_{xxxx}+6 u^3u_xu_{xxx}+4u^3u_{xx}^2+7u^2u_x^2u_{xx}\notag
\end{gather}
which is part of hierarchy $\mathbf{I}$, namely
\begin{gather}
u_t=\left(R_1^2[u]+R_1[u]C_1+C_0\right)\left(h_1u_{xx}+\{h_1\}u_x^2\right),
\notag
\end{gather}
with $h_1=u^2$ and $C_1=C_0=0$. This equation linearises to
\begin{gather}
U_T=U_{XXXX}\notag
\end{gather}
by the transformation
\begin{gather}
x=U,\quad
dt=dT,\quad
u(x,t)=U_X.\notag
\end{gather}

\noindent
{\bf Fifth-order equations:}

\strut\hfill

\noindent
An equation by S Kawamoto \cite{kawamoto}:
\begin{gather}
u_t=u^5u_{xxxxx}+10u^4u_xu_{xxxx}
+15u^4u_{xx}u_{xxx}
+25u^3u_x^2u_{xxx}\notag\\
\qquad
+30u^3u_xu_{xx}^2
+15u^2u_x^3u_{xx}\notag
\end{gather}
which is part of hierarchy $\mathbf{I}$, namely
\begin{gather}
u_t=\left(R_1^3[u]+R_1^2[u]C_2+R_1[u]C_1+C_0\right)\left(h_1u_{xx}
+\{h_1\}u_x^2\right),\notag
\end{gather}
with $h_1=u^2$ and $C_2=C_1=C_0=0$. This equation linearises to
\begin{gather}
U_T=U_{XXXXX}\notag
\end{gather}
by the transformation
\begin{gather}
x=U,\quad
dt=dT,\quad
u(x,t)=U_X.\notag
\end{gather}

\strut\hfill

\noindent
An equation by A H Bilge \cite{bilge}:
\begin{gather}
u_t=u_{xxxxx}+\beta(u_{xxxx}u_x+2u_{xxx}u_{xx})
+\beta^2\left(\frac{2}{5}u_{xxx}u_x^2+\frac{3}{5}u_{xx}^2u_x\right)\notag\\
\qquad +\frac{2}{25}\beta^3u_{xx}u_x^3+\frac{1}{625}\beta^4u_x^5
\notag
\end{gather}
which is part of hierarchy $\mathbf{VIII}$, namely
\begin{gather}
u_t=\left(R_8^3[u]+R_8^2[u]C_2+R_8[u]C_1+C_0\right)\left(u_{xx}
+h_8u_x^2\right),\notag
\end{gather}
with $h_8=\beta/5$ ($\beta$ is an arbitrary but nonzero constant) and
$C_2=C_1=C_0=0$.
This equation linearises to
\begin{gather}
U_T=U_{XXXXX}\notag
\end{gather}
by the transformation
\begin{gather}
x=U,\quad
dt=dT,\quad
u(x,t)=\frac{5}{\beta}\ln\left|U_X\right|.\notag
\end{gather}

\subsection{Higher-order potential equations and more linearisable
hierarchies}

As in the case of the second-order equations (I) - (VIII) discussed in
Section 2, the corresponding hierarchies (I) - (VIII) may also be
written in potential form,
resulting in new forms of linearisable higher-order equations. Recursion
operators are presented here that generate these hierarchies of
linearisable equations for 
(\ref{new_I}), 
(\ref{new_IIi}), (\ref{new_IV.1}), (\ref{new_IV.2}), (\ref{new_V}),
(\ref{new_VI}), (\ref{new_VII}), and (\ref{new_VIII}). First we
consider one example in detail and then 
list some other cases.

\strut\hfill

\noindent
{\bf Detailed example for hierarchy (I):}
Consider the third-order equation of hierarchy (I), i.e.,
\begin{gather}
u_t=h_3^{3/2}\left(u_{xxx}+3\frac{\ddot h_3}{\dot h_3}u_xu_{xx}
+\frac{\dddot h_3}{\dot h_3}u_x^3\right)+C \left(h_3u_{xx}+\{h_3\}_u
u_x^2\right).
\end{gather}
Let
\begin{gather}
v_x=h_3^{-1/2}(u)+\phi_1(x)\notag\\
v_t=-\frac{1}{2}\left(\dot h_3(u)u_{xx}+\ddot
h_3(u)u_x^2+Ch_3^{-1/2}\dot h_3 u_x\right)-\lambda_1.
\end{gather}
The potential equation takes the form
\begin{gather}
\label{3rd-I-Pot}
v_t=\frac{v_{xxx}-\phi_{1xx}(x)}{(v_x-\phi_1(x))^3}
-3\frac{(v_{xx}-\phi_{1x})^2}{(v_x-\phi_1(x))^4}+\frac{v_{xx}-\phi_{1x}(x)}{
(v_x-\phi_1(x))^2}-\lambda_1.
\end{gather}
Transforming (\ref{3rd-I-Pot}) by the pure hodograph transformation
\begin{gather}
v(x,t)=\chi,\quad
t=\tau,\quad
x=V(\chi,\tau)
\end{gather}
leads to the autonomous equation
\begin{gather}
\label{3rd-I-New}
V_\tau=
\frac{V_{\chi\chi\chi}+\phi_1''V_\chi^4}{(1-\phi_1V_\chi)^3}
+3\,\frac{\phi_1V_{\chi\chi}^2+2\phi_1'V_\chi^2V_{\chi\chi}
+(\phi_1')^2V_\chi^5}{(1-\phi_1V_\chi)^4}
+C\frac{V_{\chi\chi}+\phi_1' V_\chi^3}{(1-\phi_1V_\chi)^2}
+\lambda_1 V_\chi,
\end{gather}
where $\phi_1'=d\phi_1(V)/dV$. Equation (\ref{3rd-I-New})
linearises to
\begin{gather}
U_T=U_{XXX}+CU_{XX}+\lambda_1 U_X\notag
\end{gather}
by the transformation
\begin{gather}
V(\chi,\tau)=U(X,T),\quad
d\tau=dT,\quad
V_\chi^{-1}-\phi_1(V)=U_X^{-1}.\notag
\end{gather}
This is transformation (\ref{trans-I-new}), i.e.,
the same transformation that linearises (\ref{new_I}), that is
\begin{gather}
V_\tau=\frac{V_{\chi\chi}+\phi_1'(V)V_\chi^3}{(1-\phi_1(V)V_\chi)^2}+\lambda_1
V_\chi.\notag
\end{gather}
Calculating the recursion operator for (\ref{new_I}) we obtain
\begin{gather}
\label{tilde-R_1}
\tilde R_1[V]=\frac{1}{1-\phi_1(V)V_\chi}D_\chi
+\frac{\phi_1(V)V_{\chi\chi}+\phi_1'(V)V_\chi^2}
{(1-\phi_1(V)V_\chi)^2},
\end{gather}
so that
\begin{gather}
V_\tau=\left(\tilde R_1^2[V]+C\tilde R_1[V]+\lambda_1\right)V_\chi
\end{gather}
leads to (\ref{3rd-I-New}). A new hierarchy of linearisable
equations is therefore given by
\begin{gather}
\label{hier-I}
V_\tau=\left(\tilde R_1^m[V]+\sum_{i=0}^{m-1}\tilde
R_1^i[V]C_i\right)V_\chi.
\end{gather}
The hierarchy (\ref{hier-I}) is linearised by the transformation
(\ref{trans-I-new}). 


\strut\hfill

We list below the recursion operators for equations
(\ref{new_IIi}), (\ref{new_IV.1}), (\ref{new_IV.2}), (\ref{new_V}),
(\ref{new_VI}), (\ref{new_VII}), and (\ref{new_VIII}).

\strut\hfill

\noindent
{\bf Equation (\ref{new_IIi}) of (II):} The recursion operator is
\begin{gather}
\label{tilde-R_2}
\tilde R_2[V]=\frac{1}{1-\phi_2(V)V_\chi}D_\chi
+\frac{\phi_2(V)V_{\chi\chi}+\phi_2'(V)V_\chi^2}{(1-\phi_2(V)V_\chi)^2}
+\lambda \frac{V_\chi}{1-\phi_2(V)V_\chi}
\end{gather}
and the hierarchy
\begin{gather}
\label{hier-II}
V_\tau=\left(\tilde R_2^m[V]+\sum_{i=0}^{m-1}\tilde
R_2^i[V]C_i\right)V_\chi\notag
\end{gather}
is linearisable by the transformation (\ref{trans-II-new}).

\strut\hfill

\noindent
{\bf Equation (\ref{new_IV.1}) of (IV.1):} The recursion operator is
\begin{gather}
\label{tilde-R_4.1}
\tilde R_{4.1}[V]=\frac{1}{V_\chi}D_\chi-\frac{V_{\chi\chi}}{V_\chi^2}
-r
\end{gather}
and the hierarchy
\begin{gather}
V_\tau=\left(\tilde R_{4.1}^m[V]+\sum_{i=0}^{m-1}\tilde
R_{4.1}^i[V]C_i\right)
\left(\frac{V_{\chi\chi}}{V_\chi^2}+\phi_4'(V)V_\chi
+(2r-\lambda_1)(1-\phi_4(V)V_\chi)\right)
\end{gather}
is linearisable by the transformation (\ref{trans-IV.1-new}).
Here $r$ is given by (\ref{r}).

\strut\hfill

\noindent
{\bf Equation (\ref{new_IV.2}) of (IV.2):} The recursion operator is
\begin{gather}
\label{tilde-R_4.2}
\tilde R_{4.2}[V]=\frac{1}{V_\chi}D_\chi-\frac{V_{\chi\chi}}{V_\chi^2}
-\lambda_1
\end{gather}
and the hierarchy
\begin{gather}
V_\tau=\left(\tilde R_{4.2}^m[V]+\sum_{i=0}^{m-1}\tilde
R_{4.2}^i[V]C_i\right)
\left(\frac{V_{\chi\chi}}{V_\chi^2}+\phi_4'(V)V_\chi
+\lambda_1(1-\phi_4(V)V_\chi)-\frac{1}{\lambda_1}e^{\lambda_1 V}V_\chi\right)
\end{gather}
is linearisable by the transformation (\ref{trans-IV.2-new}).

\strut\hfill

\noindent
{\bf Equation (\ref{new_V}) of (V):} The recursion operator has the same form
as (\ref{tilde-R_2}), namely
\begin{gather}
\label{tilde-R_5}
\tilde R_5[V]=\frac{1}{1-\phi_5(V)V_\chi}D_\chi
+\frac{\phi_5(V)V_{\chi\chi}+\phi_5'(V)V_\chi^2}{(1-\phi_5(V)V_\chi)^2}
+\lambda \frac{V_\chi}{1-\phi_5(V)V_\chi}
\end{gather}
and the hierarchy 
\begin{gather}
\label{hier-V}
V_\tau=\left(\tilde R_5^m[V]+\sum_{i=0}^{m-1}\tilde
R_5^i[V]C_i\right)
\left(\frac{V_{\chi\chi}+\phi_5'(V)V_\chi^3}{(1
-\phi_5(V)V_\chi)^2}+\frac{\lambda V_\chi^2}{1-\phi_5 V_\chi}
+\frac{\lambda_2}{\lambda}(1-\phi_5(V)V_\chi)\right)\notag
\end{gather}
is linearisable by the transformation (\ref{trans-V-new}).

\strut\hfill

\noindent
{\bf Equation (\ref{new_VI}) of (VI):} The recursion operator is
\begin{gather}
\label{tilde-R_6}
\tilde R_6[V]=\frac{1}{V_\chi}D_\chi-\frac{V_{\chi\chi}}{V_\chi^2}
+\frac{1}{2}\left(\frac{1-\phi_6V_\chi}{V_\chi}\right)
\end{gather}
and the hierarchy 
\begin{gather}
V_\tau=\left(\tilde R_6^m[V]+\sum_{i=0}^{m-1}\tilde
R_6^i[V]C_i\right)
\left(\frac{V_{\chi\chi}}{V_\chi^2}+\phi_6'(V)V_\chi
-\frac{1}{2}\frac{(1-\phi_6(V)V_\chi)^2}{V_\chi}\right),
\end{gather}
is linearisable by the transformation (\ref{trans-VI-new}).

\strut\hfill

\noindent
{\bf Equation (\ref{new_VII}) of (VII):} The recursion operator has
the same form as (\ref{tilde-R_1}), namely
\begin{gather}
\label{tilde-R_7}
\tilde R_7[V]=\frac{1}{1-\phi_7(V)V_\chi}D_\chi
+\frac{\phi_7(V)V_{\chi\chi}+\phi_7'(V)V_\chi^2}{(1-\phi_7(V)V_\chi)^2}
\end{gather}
and the hierarchy 
\begin{gather}
\label{hier-VII}
V_\tau=\left(\tilde R_7^m[V]+\sum_{i=0}^{m-1}\tilde
R_7^i[V]C_i\right)
\left(\frac{V_{\chi\chi}+\phi_7'(V)V_\chi^3}{(1
-\phi_7(V)V_\chi)^2}
+\lambda_3(1-\phi_7(V)V_\chi)\right)\notag
\end{gather}
is linearisable by the transformation (\ref{trans-VII-new}).

\strut\hfill

\noindent
{\bf Equation (\ref{new_VIII}) of (VIII):} The recursion operator is
of the same form as (\ref{tilde-R_4.2}), namely
\begin{gather}
\label{tilde-R_8}
\tilde R_8[V]=\frac{1}{V_\chi}D_\chi-\frac{V_{\chi\chi}}{V_\chi^2}
-\lambda_1
\end{gather}
and the hierarchy 
\begin{gather}
V_\tau=\left(\tilde R_8^m[V]+\sum_{i=0}^{m-1}\tilde
R_8^i[V]C_i\right)
\left(\frac{V_{\chi\chi}}{V_\chi^2}+\phi_8'(V)V_\chi
+\lambda_1(1-\phi_8(V)V_\chi)\right)
\end{gather}
is linearisable by the transformation (\ref{trans-VIII-new}).

\strut\hfill

\noindent
{\it Remark: For (\ref{new_IIIi}) and (\ref{new_IIIii}) of {\bf (III)}
we were not able to find recursion operators. The operators are not of
a similar form
as those given above.}

\subsection{Autohodograph transformations}

In \cite{eul_gan_eul_lin} we introduced an autohodograph transformation
for the hierarchy
\begin{gather}
\label{gand}
u_t=R^m[u](u^{-2}u_x)_x,
\end{gather}
where
\begin{gather}
R[u]=D_x^2u^{-1}D_x^{-1}\equiv u^{-1}D_x-2u^{-2}u_x
-(u^{-2}u_{xx}-2u^{-3}u_x^2)D^{-1}_x .
\end{gather}
An autohodograph transformation is an $x$-generalised hodograph
transformation that keeps the equation invariant. For the
hierarchy (\ref{gand})
the autohodograph tranformation is
\beg
\left\{\ba{l}
\displaystyle{dX(x,t)
=xdx
+\left\{xD_{x}^{-1}R^{m}[u]\left(u^{-2}u_x\right)_x
+\left(u^{-1}D_x\right)^mu^{-1}\right\}dt}
\\[3mm]
dT(x,t)=dt \\[3mm]
U(X,T)=x^{-1},
\ea\right.
\eeq
The hierarchy (\ref{gand}) is a special case of hierarchy (I),
with $h_1(u)=u^{-2}$. The result given in \cite{eul_gan_eul_lin}
can be generalised
to the hierarchy (I), with arbitrary $h_1$ and $R_1$ given in
subsection 3.1:
The hierarchy of evolution equations 
\begin{gather}
\label{auto-hier}
u_{t}=R_1^m[u]\left(h_1u_{xx}+\{h_1\}_{u}u_x^2\right),\quad
m=1,2,3,\ldots,
\end{gather}
admits the autohodograph transformation
\beg
\left\{\ba{l}
\displaystyle{dX_0(x,t)
=xh_1^{-1/2}dx
+\left[-\frac{1}{2}xD_{x}^{-1}h_1^{-3/2}\dot h_1
R_1^{m}[u]\left(h_1u_{xx}+\{h_1\}_{u}u_{x}^2\right)\right.}\\[3mm]
\hphantom{T(x,t)=+}
\displaystyle{\left.
+\frac{1}{2}D_{x}^{-2}h_1^{-3/2}\dot h_1
R_1^{m}[u]\left(h_1u_{xx}+\{h_1\}_{u}u_{x}^2\right)\right]
dt}
\\[3mm]
dT_0(x,t)=dt \\[3mm]
U_0(x_0,t_0)=x^{-1},
\ea\right.
\eeq
thereby transforming (\ref{auto-hier}) to
\begin{gather}
U_{0T_0}=R_1^{m}[U_0]\left(h_1U_{0X_0X_0}+\{h_1\}_{U_0}U_{0X_0}^2\right),\quad
m=1,2,3,\ldots\ .
\end{gather}

\subsection{On the $x$-dependent linear equation}

In our classification of equations (I) - (VIII) we use as starting
point the autonomous linear equation (\ref{lin_auto2}). Alternatively
we could allow the $\lambda$'s to be $x$-dependent continuous
functions. In such case one applies the pure hodograph
transformation
followed by the $x$-generalised hodograph transformation, in order to
construct linearisable autonomous evolution equations.
The same result then follows, i.e., equations (I) - (VIII), with
only one more additional autonomous equation, namely the equation
obtained from the linear equation after transforming by the
pure hodograph transformation.

Consider
\begin{gather}
U_T=\lambda_0(X)+\sum_{k=1}^n\lambda_k(X) U_{X^k}.
\end{gather}
Transforming the equation by the pure hodograph transformation
\begin{gather}
X=\omega(\xi,\eta),\quad U(X,T)=\xi,\quad T=\eta,\notag
\end{gather}
leads to the nonlinear evolution equation
\begin{gather}
\label{x-dep-nonlin}
\omega_\eta=-\lambda_0(\omega)\omega_\xi
-\sum_{j=0}^n \lambda_{j+1}(\omega)\,\left(D_\xi
\omega_\xi^{-1}\right)^j,
\end{gather}
that is
\begin{gather}
\omega_\eta =-\lambda_0(\omega)\omega_\xi
-\lambda_1(\omega)
+\lambda_2(\omega)\omega_\xi^{-2}\omega_{\xi\xi}
-\lambda_3(\omega)\left(3\omega_\xi^{-4}\omega_{\xi\xi}^2
-\omega_\xi^{-4}\omega_{\xi\xi\xi}\right)
-\cdots\ .\notag
\end{gather}
Using the $x$-generalised hodograph transformation 
to transform (\ref{x-dep-nonlin}) leads to the same
hierarchies (I) - (VIII),
besides classes of explicitly $x$-dependent evolution
equations.

\section*{Acknowledgements}

ME acknowledges financial support from the {\it Knut and Alice Wallenberg
Foundation} under grant Dnr. KAW 2000.0048.

\label{euler-lastpage}


\begin{thebibliography}{20}
\small
\bibitem{bilge}
Bilge A H, On the Equivalence of Linearization and Formal Symmetries
as Integrability Tests for Evolution Equations,
{\it J. Phys. A: Math. Gen.} {\bf 26} (1993), 7511 -- 7519.

\bibitem{Bluman}
Bluman G W and Kumei S, Symmetries and Differential Equations,
AMS {\bf 81}, Springer-Verlag, New York, 1989.

\bibitem{calogero91}
Calogero F, Why are Certain Nonlinear PDEs Both Widely Applicable
and Integrable?, in What Is Integrability?, Zakharov V E (Editor),
Springer Series in Nonlinear Dynamics, Berlin, 1 -- 62, 1991.


\bibitem{clark}
Clarkson P A, Fokas A S and Ablowitz M J, Hodograph Transformations of
Linearizable Partial Differential Equations,
{\it SIAM J. Appl. Math.} {\bf 49} (1989),
1188 -- 1209.

\bibitem{eul_eul}
Euler N and Euler M, A Tree of Linearisable Second-Order
Evolution Equations by Generalised Hodograph Transformations,
{\it J. Nonlin. Math. Phys.} {\bf 8} (2001),  342-362.

\bibitem{eul_gan_eul_lin}
Euler N, Gandarias M L, Euler M and Lindblom O,
Auto-hodograph Transformations for a Hierarchy of Nonlinear Evolution
Equations, {\it J. Math. Anal. Appl.} {\bf 257} (2001), 21--28.

\bibitem{fokas80}
Fokas A S, A Symmetry Approach to Exactly Solvable Evolution
Equations,
{\it J. Math. Phys.} {\bf 21} (1980), 1318--1325.

\bibitem{fokas87}
Fokas A S, Symmetries and Integrability,
{\it Stud. Appl. Math.} {\bf 77} (1987), 253--299.

\bibitem{kawamoto}
Kawamoto S, An Exact Transformation for the Harry Dym Equation to the
Modified KdV Equation,
{\it J. Phys. Soc. Japan} {\bf 54} (1985) 2055--2056.

\bibitem{Olver77}
Olver P J, Evolution Equations Possessing Infinitely Many Symmetries,
{\it J. Math. Phys.} {\bf 18} (1977), 1212--1215.

\bibitem{Olver}
Olver P J, Applications of Lie Groups to Differential Equations, GTM {\bf 107},
Springer-Verlag, New York, 1986.



\end{thebibliography}
\end{document}